\documentclass[useAMS,usenatbib]{mn2e}

\usepackage{epsf}
\usepackage{amsmath}

\newcommand{\kmsmpc}{\kms\;{\rm Mpc}^{-1}}
\newcommand{\lya}{Ly$\alpha$\ }
\newcommand{\hkpc}{h^{-1}{\rm kpc}}
\newcommand{\hmpc}{h^{-1}{\rm Mpc}}

\newcommand{\kms}{\;{\rm km}\,{\rm s}^{-1}}
\newcommand\cdunits{{\rm cm}^{-2}}

\newcommand{\gad}{{\sc Gadget-2}}
\newcommand{\ion}[2]{\hbox{#1\,{\sc #2}}}

\title[The Enrichment History of Baryons]{The Enrichment History of Baryons in the Universe}

\author[R. Dav\'e \& B. D. Oppenheimer]{Romeel Dav\'e \& Benjamin D. Oppenheimer\\Astronomy Department, University of Arizona, Tucson, AZ 85721}

\begin{document}

\pubyear{2006}

\maketitle

\label{firstpage}

 \begin{abstract}
We present predictions for the cosmic metal budget in various phases
of baryons from redshift $z=6\rightarrow 0$, taken from a cosmological
hydrodynamic simulation that includes a well-constrained model for
enriched galactic outflows.  We find that substantial amounts of
metals are found in every baryonic phase at all epochs, with diffuse
intergalactic gas dominating the metal budget at early epochs and stars
and halo gas dominating at recent epochs.  We provide a full accounting
of metals in the context of the missing metals problem at $z\approx
2.5$, showing that $\sim40$\% of the metals are in galaxies, and the
remainder is divided between diffuse IGM gas and shocked gas in halos
and filamentary structures.  Comparisons with available observations
of metallicity and metal mass fraction evolution show broad agreement.
We predict stars have a mean metallicity of one-tenth solar already
at $z=6$, which increases slowly to one-half solar today, while stars
just forming today have typically solar metallicity.  Our \ion{H}{i}
column density-weighted mean metallicity (comparable to Damped \lya\
system metallicities) slowly increases from one-tenth to one-third solar
from $z=6\rightarrow 1$, then falls to one-quarter solar at $z=0$.
The global mean metallicity of the universe tracks $\sim 50$\% higher
than that of the diffuse phase down to $z\sim 1$, and by $z=0$ it has
a value around one-tenth solar.  Metals move towards higher densities
and temperatures with time, peaking around the mean cosmic density at
$z=2$ and an overdensity of 100 at $z=0$.  We study how carbon and
oxygen ions trace the path of metals in phase space, and show that
\ion{O}{iii}-\ion{O}{vii} lines provide the most practical option for
constraining intergalactic medium metals at $z\la 2$.
\end{abstract}

 
\section{Introduction}

Heavy elements are produced exclusively in stars, and then distributed
throughout the Universe through supernovae and stellar winds.  As such,
they provide a unique tracer of the star formation and feedback processes
that are central to understanding how galaxies form and evolve.
Metals are present at the earliest observable epochs: The farthest
known quasar at $z=6.4$ shows substantial enrichment despite being less
than 1~Gyr old~\citep{bar03}, and metals are seen in the intergalactic
medium (IGM) at $z\ga 6$~\citep{bec06,rya06,sim06}.  At lower redshifts,
the galaxy mass-metallicity relation has been quantified out to $\sim
2$~\citep{tre04,sav05,erb06}, intracluster gas metallicities have been
constrained out to $z\sim 1$~\citep{ros04} and the IGM is observed in
metal lines nearly continuously from $z\sim 6\rightarrow 0$.

\citet{pet99} first noted the interesting fact that the metals seen
in Lyman break galaxies at $z\sim 2-3$ fall well short of the amount
expected to be produced by the stars in those galaxies.  This is known
as the {\it missing metals problem}.  It implies that most metals
must either reside outside of galaxies, or be in some non-observable
form in or around galaxies.  Attempts to quantify metals in the IGM
\citep[][hereafter FSB05]{fer05} or other galaxy populations such as
sub-millimeter galaxies~\citep{bou06a} or damped Lyman alpha (DLA)
systems~\citep{pro06} have not clearly been able to account for the
shortfall, though uncertainties remain substantial~\citep[hereafter
BLP06]{bou06b}.  Virtually every galaxy class and environment examined
seems to contain a small but non-trivial amount of metals, so it appears
that metal pollution has been curiously democratic.  As the distribution
of metals in various environments and at various epochs becomes better
measured, this provides important constraints into galactic feedback
processes.

A theoretical framework for understanding these wide ranging observations
is provided by simulations of galaxy formation that include metal
production and distribution mechanisms.  The pioneering work of
\citet{cen99b} using cosmological hydrodynamic simulations highlighted
some basic trends arising in such models, such as the early enrichment
of galaxies and the natural establishment of a metallicity gradient
with cosmic overdensity.  Using more advanced, higher-resolution
cosmological simulations, \citet{tha02} studied global metal enrichment
from dwarf galaxies down to $z=4$, and showed that it is possible to
enrich the IGM at early times to a value broadly consistent with data
while producing dwarf galaxies with enrichment levels similar to local
dwarfs.  

An alternative approach was taken by \citet{cal04}, who used detailed
chemical evolution models for various galaxy types together with
observations of luminosity functions in those types to construct
the evolution of metals in galaxies, intracluster gas, and IGM gas.
While their models make detailed predictions for individual elements in
various galaxy classes, they do not embed the predictions within an ab
initio hierarchical structure formation framework, relying instead on
passively evolving observed luminosity functions back in time.  Still,
this provides some interesting predictions that are not dissimilar to
those presented here.  In \citet{cal06} they extend their approach to
study IGM enrichment, and suggest that winds with velocities exceeding
1000~km/s are required to pollute the IGM to observed levels.

In \citet[][hereafter OD06]{opp06} we took such studies to the next
level of sophistication, using cosmological simulations that incorporate
enriched kinetic feedback in a full hierarchical structure formation
setting.  Our models, employing outflow scaling relations taken from local
starburst data, showed remarkable success at reproducing observations
of \ion{C}{iv} enrichment in the IGM from $z\sim 6\rightarrow 2$.
Matching these data generally required highly mass loaded but low
velocity winds from early galaxies, so as to eject a substantial metal
mass into the IGM without overheating it.  Such scalings naturally arise
in momentum-driven wind scenarios~\citep{mur05} that are also favored by
local outflow observations~\citep{mar05,rup05}, providing an intriguing
connection between outflows at all epochs.  In \citet{dav06} we show
that significant early feedback is also necessary in order to suppress
reionization-epoch star formation.  In Finlator, Dav\'e \& Oppenheimer
(in preparation) we show that the resulting galaxy mass-metallicity
relation from our favored outflow models agrees well with observations
at $z\sim 2$.  These successes suggest that we now have a plausible
structure formation framework with which to study the global evolution
of metals in the Universe.

In this paper we use cosmological hydrodynamic simulations
incorporating our most successful feedback model from OD06 (described in
\S\ref{sec:sims}) to study the evolution of metals in various baryonic
phases in the Universe from $z=6\rightarrow 0$ (\S\ref{sec:metphase})
and provide quantitative answers for the missing metals problem
(\S\ref{sec:missing}).  We go on to compare the metallicities in
those phases with observations (\S\ref{sec:metallicity}), showing
broad agreement.  We make predictions for how metals evolve in cosmic
phase space (\S\ref{sec:phasespace}), and study how different oxygen
and carbon ions may be used to probe various regimes of phase space
(\S\ref{sec:tracers}).  We present our conclusions in \S\ref{sec:summary}.

\section{Metal Evolution}\label{sec:metevol}

\subsection{Simulations}\label{sec:sims}

We run a cosmological hydrodynamic simulation using
\gad~\citep{spr02,spr03a} with improvements as described in OD06.
Our run has $2\times 256^3$ particles in a randomly-generated volume of
$(32\hmpc)^3$, with an equivalent Plummer softening length of $2.5\hkpc$
(comoving).  Initial conditions were generated using an \citet{eis99}
power spectrum, assuming a concordance cosmology~\citep{spe03} having
$\Omega=0.3$, $\Lambda=0.7$, $n=1$, $H_0=70\kmsmpc$, and $\sigma_8=0.9$.
The simulation was started at $z=159$, in the linear regime, and evolved
to $z=0$.  We include metal-line cooling by augmenting the primordial
cooling rate calculated as described in \citet{kat96} with the difference
in cooling rates between the primordial and metal-enriched collisional
ionization equilibrium models of \citet{sut93}.  Star formation is
modeled in a multi-phase manner as described in~\citet{spr03a}, occuring
in regions sufficiently dense to undergo multi-phase collapse, with the
threshold density depending on the metallicity through the cooling rate.

We employ a momentum-driven wind model as described in OD06, in particular
the vzw model, which provides the overall best match to the \ion{C}{iv}
observations considered.  In our heuristic outflow model, a star forming
particle has a probability to enter a wind that is proportional to the
assumed mass loading factor, and if selected it is given a velocity
kick.  In the vzw prescription, the wind velocity is proportional to
the galaxy velocity dispersion, which we estimate from the local value
of the gravitational potential.  The value of the velocity kick has
some spread associated with an observed spread in the the luminosity
factor~\citep{rup05}; see \citet{opp06} for full formula.  As predicted
in a momentum-driven wind model~\citep[e.g.][]{mur05}, we assume that
the mass loading factor scales inversely with the velocity dispersion.

We identify galaxies using Spline Kernel Interpolative Denmax (SKID),
and halos using a spherical overdensity algorithm, as described
in \citet{fin06}.  Our simulation galaxy mass resolution limit is
$M_*>9.91\times 10^8 M_\odot$ (i.e. 64 star particles); all predictions
regarding galaxies are implicitly for galaxies above this stellar mass.

Our simulations assume a constant metal yield of 0.02, hence metal
production directly tracks star formation.  Others prefer a value of 1/42
to relate star formation to metal mass production~\citep[e.g.][]{mad96}.
Most of the results presented here will not depend sensitively on this
choice, and in fact as long as this value is constant with time, the {\it
fractions} of metals in various phases will be essentially unaffected
(modulo small differences due to metal-line cooling).  We assume a solar
metallicity value of 2\% metals by mass.

While we focus on the vzw model here, the mzw model (the other
momentum-driven wind model favored by OD06) produces similar results.
We choose the $32\hmpc$ vzw run for this work because it has sufficient
resolution to capture early star formation, while still having enough
volume to be meaningfully representative at $z=0$.  On a practical note,
it is the only model we have evolved to $z=0$, owing to the significant
computational expense involved.  We note that cosmic variance could still
be significant, and the wind model is as yet not tightly constrained,
so the exact values quoted here should be taken as illustrative rather
than precise predictions.  Using a $(16\hmpc)^3$ volume vzw run, we
have checked that resolution effects introduce only minor deviations
from the numbers quoted herein from $z=6\rightarrow 1.5$.

Although we are less confident of our predictions at $z\la 2$ because
this particular feedback model has not been critically tested there,
we include predictions anyway to illustrate some basic physical trends
that are likely to be valid.  As we will show later, the comparisons to
data at $z\la 2$ seem to indicate broad agreement.

\subsection{Metal Mass Fractions}\label{sec:metphase}

\begin{figure}
\vskip -0.5in
\setlength{\epsfxsize}{0.6\textwidth}
\centerline{\epsfbox{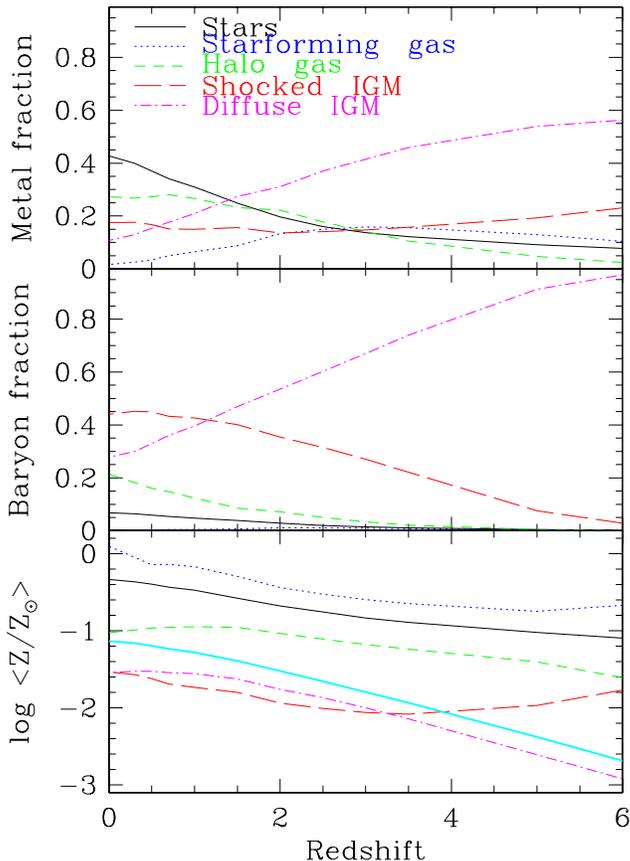}}
\vskip -0.5in
\caption{{\it Top:} Metal mass fraction in various baryonic phases as a 
function of redshift: Stars (black solid), 
star-forming gas (blue dotted), halo gas (green dashed), shocked 
IGM ($T>10^{4.5}K$; red long-dashed), and diffuse IGM (cyan dot-dashed).
{\it Middle:} As above, for total mass fraction of baryons.
{\it Bottom:} As above, for metallicity.  The thick cyan line represents
the mean mass-weighted metallicity of the universe.
}
\label{fig:metphase}
\end{figure}

Figure~\ref{fig:metphase} (top panel) shows the evolution of the metal
mass fraction in various baryonic phases:  Stars, star-forming gas,
halo gas, shocked IGM, and diffuse IGM.  Middle panel shows the mass
fraction in baryons for comparison.  Star-forming gas corresponds to
gas that is above the density threshold for multi-phase fragmentation in
the \citet{spr03a} recipe; this may be thought of as insterstellar medium
(ISM) gas.  Halo gas is defined as gas that is within identified halos but
is not star-forming; this includes e.g. hot intracluster gas.  Shocked and
diffuse IGM gas encompass all gas outside of halos, and are separated
by a temperature threshold of $30,000$~K~\citep{dav99}.  We choose this
threshold as opposed to the canonical warm-hot intergalactic medium
~\citep[WHIM;][]{cen99,dav01} threshold of $T>10^5$~K because in our
models, substantial amounts of \ion{C}{iv} absorption comes from gas right
around $10^5$~K (OD06), whose temperature is set not by photoionization
but by shock heating from outflows and/or gravitational collapse.
Hence this gas rightfully belongs in the shocked IGM component.

From $z\sim 6\rightarrow 1.5$, diffuse gas is the largest reservoir of
metals; below this, stars take over with halo gas not far behind.  It may
be surprising that at early times the diffuse phase harbors most metals,
given that metals are produced in much denser regions.  This arises
because early blowout is quite effective in our models, which is required
to both enrich the IGM early on and suppress early star formation to match
observed high-redshift luminosity functions~\citep{dav06}.  The diffuse
IGM contains virtually all baryons at high redshift as is generically
seen in hydrodynamic simulations~\citep[e.g.][]{dav01}, then drops steadily to
around 30\% mass fraction by today.  The drop in metal fraction tracks
the drop in total baryon fraction in this phase.

As galaxies grow and outflows become less effective at polluting the
diffuse IGM in our momentum-driven wind scenario, the metal mass fraction
that remains locked in stars increases.  By $z\sim 2$, stars contain
$20\%$ of all metals, and by today they hold over 40\%.  This value
is consistent within uncertainties to the value of 52\% predicted in
the model of \citet{cal04}.  Note that the stellar baryon fraction
at $z=0$ is $6.8\%$, which is in agreement with observational
estimates~\citep[e.g.][]{bel03}.

The amount of metals in shocked IGM gas remains fairly stable at $15-20\%$
at all epochs.  The mass fraction of baryons in this phase grows almost
linearly with redshift up to $45\%$ today.  Hence while the shocked
IGM contains a significant fraction of all present-day baryons, it
does not hold a commensurately large reservoir of metals.  Meanwhile,
the growth of larger potential wells and reaccretion of IGM metals
into halos results in a gradual increase of metal mass in halo gas,
up to around one-quarter of all metals today.  Halo gas is generally
over-enriched relative to its mass fraction.

Star-forming gas shows a mild increase in metal content at early times,
but after $z=3$ it drops steadily to a very small number at the present
epoch.  This traces its baryon fraction evolution, which is difficult
to see on this plot because it is quite small at all epochs.  This shows
that galaxies are in general becoming more gas-poor with time, as inflow
shuts off because of cosmological acceleration combined with the growth
of large potential wells.

In summary, a plausible model of cosmic enrichment that broadly
matches key IGM and galaxy metallicity measures predicts that metals are
present in a range of baryonic phases from the earliest epochs until the
present day.  There is no dominant reservoir of metals at any epoch,
although diffuse IGM gas holds a significant amount early on (mainly
because virtually all baryons are in this phase), while stars hold the
most metals today. 

\subsection{Missing Metals Problem}\label{sec:missing}

Observational estimates have now been made for the metal content in
various systems at $z\sim 2.5$, which as discussed earlier has led to
a missing metals problem.  Here we present a comparison of our model
predictions with various observations that try to account for the
missing metals.

At $z\sim 2-3$, FSB05 and BLP06 estimated that stars in Lyman break
galaxies, particularly the BX sample of \citet{sha05}, contain about 18\%
of all metals.  BLP06 go on to estimate a contribution of $\sim
5\%$ from so-called Distant Red Galaxies, though this may have some
overlap with the BX sample~\citep{red05}.  Our predicted metal fraction
in stars is somewhat higher ($\approx 30\%$), which may be explained if
our typical stellar mass of LBGs is higher by $\sim 50$\% than derived
by \citet{sha05}.  This would easily fall within the substantial
uncertainties of spectral energy distribution fitting techniques by
which stellar masses are estimated.  In general, simulations tend to
predict LBGs that have substantial older populations~\citep{fin06},
and tend to be more massive than burst-dominated models would predict.
Furthermore, there are completeness issues as our simulations include
all stars in galaxies down to $M_*\sim 10^9M_\odot$, while typical LBG
surveys are substantially incomplete at such masses.  

At $z\sim 0$, an observational census of metals by \citet{pag02}
finds $\sim 50$\% in stars today depending on assumptions about the
IMF and stellar evolution, which again is in general agreement with
our prediction of 43\%.  If we fix a stellar baryon fraction of 7\% as
our models predict, then \citet{fin03} determines $\Omega_{Z,*}\approx
3\times 10^{-5}$, whereas we predict $2.5\times 10^{-5}$, again within
uncertainties.  In all, while more careful comparisons are warranted, the
preliminary indication is that our stellar metal fraction agrees broadly
with data, and shows that stars contain a substantial but not dominant
fraction of metals in the universe at both high and low redshifts.

Our simulations further predict that about 15\% of metals reside
in cold star-forming gas in galaxies at $z=2.5$.  It has been
suggested that Damped Lyman Alpha absorbers (DLAs) may trace this
cold dense gas.  Yet despite the fact that such systems dominate
the \ion{H}{i} content of the Universe, they are seen to contain
just a few percent of all metals~\citep{pro06}.  However, DLAs may
just be the tip of the iceberg: \citet{pro06} and \citet{kul06} find
sub-DLAs ($10^{19}<N_{HI}<10^{20.3}\cdunits$ systems) with above-solar
metallicities, and extrapolate that such systems may represent $15-25$\%
or more of the total metal budget.  However, there is a bias towards
selecting highly enriched sub-DLAs, so the extrapolations are highly
uncertain.  We will address sub-DLA metallicities in the next section.
Another reservoir of star-forming gas is sub-millimeter galaxies, which
are rare but massive and highly enriched; from their gas content and
metallicity \citet{bou06a} estimated that their ISM may contain $\sim
5\%$ of the total metals.  In general, it seems that the metals in
star-forming gas are a relatively small fraction of the total metals,
broadly agreeing with our predictions, although if sub-DLAs make as much
of a contribution as claimed then perhaps the agreement is not so good.

Combining these constraints, BLP06 determined that approximately 30\%
of metals are currently observed in galaxies (stars+ISM) at $z=2.2$,
which could plausibly increase to 50\% with incompleteness corrections.
This is quite consistent with our prediction of $\sim 40\%$ of metals
being contained in galaxies with $M_*\ga 10^9M_\odot$ at these epochs.
In particular, we do not predict, and our models do not require, a
substantial amount of metals to be in some hidden form within galaxies.

The remaining 60\% of metals are not in the inner regions of galaxies,
so have presumably been ejected through galactic outflows (note that the
amount actually ejected is much larger, but many metals get reaccreted
into halos).  Of this, we predict that more than half of that is in the
diffuse IGM at $z\sim 3$.  This is substantially larger than previous
estimates.  Using two-phase analytic modeling of the IGM, FSB05 estimate
that less than 10\% of metals are allowed in this diffuse phase \citep[see
also][and references therein]{pet04}.  The discrepancy arises because the
ionization corrections needed to obtain the metallicity are typically
done assuming the metals lie in gas having the density-temperature
relation of the \lya\ forest.  However, our simulations predict that a
substantial amount of \ion{C}{iv}-absorbing gas is heated to somewhat
higher temperatures~(see Figure 9 of OD06).  Therefore we can hide
more \ion{C}{iv} in the diffuse phase (defined here as $T<10^{4.5}$~K)
because our ionization fractions are lower than given by the \lya\
equation of state (see Figure~10 of OD06).  Note that we still obtain
a good fit to the observed $b$-parameter and \ion{C}{iii}/\ion{C}{iv}
pixel optical depth ratio data, both measures of \ion{C}{iv} absorber
temperatures, despite this mildly shock-heated absorbing gas.

The remaining 30\% of metals in the Universe are divided between shocked
IGM gas and halo gas at $z\sim 2.5$.  FSB05 suggest that up to 90\% of
all metals may be hidden in hot galactic halo gas; our simulations do
not support this idea.  The amount in halo gas is difficult to quantify
at high redshifts, but an estimate of metals in clusters at $z\sim 2.5$
by \citet{fin03} predicts around 10\%, rising modestly to $20-25$\%
at $z\sim 0$.  Our predicted value for halo gas also evolves slowly as
well, and the numbers are similar, although a more careful comparison
is warranted in order to critically assess how our definition of halo
gas relates to gas in clusters or protoclusters.

The fraction of metals in shocked IGM gas, closely related to the
missing baryons~\citep{cen99} or WHIM gas, is relatively constant.
Such gas may be traced by \ion{O}{vi} absorbers~\citep{tri06}, though
because \ion{O}{vi} is not the dominant ionization state of this gas (as
we will show in \S\ref{sec:tracers}), the uncertainties in the ionization
corrections are large~\citep{tri01}.  Estimates of the metal content of
\ion{O}{vi} systems by \citet{fin03} also show little evolution in their
metal fraction, though they estimate around one-third of all metals are in
this phase as opposed to our predicted values of $15-20$\%.

So let us return to the original question: Where are the missing metals?
According to BLP06 and our models, at least 50\% are not in galaxies.
The basic answer from our simulations is that they are mostly in gas that
has been mildly heated by a combination of gravitational and feedback
effects.  The enriched diffuse IGM is somewhat warmer than expected
from photoionization alone, the WHIM at these redshifts contains a
non-trivial amount, and galactic halos also contain hot gas that has
a non-negligible metal content.  Hence metals are hidden in hot gas,
but that gas is not confined to halos but rather spread among a wide
range of overdensities (as we will show in \S\ref{sec:phasespace}).
Overall, the simulations support a view where metals are rather evenly
distributed among the various baryonic phases, which seems to be what
observations suggest as well.

\subsection{Metallicity Evolution}\label{sec:metallicity}

While obtaining a census of total metals is difficult because it
requires detecting all metals in a given phase or extrapolating
from available data, determining the mean metallicity is somewhat easier
as it may have a characteristic value for a given phase of baryons.
Hence the observations are somewhat better constrained, and in turn
provide interesting constraints on our global enrichment model.

The bottom panel of Figure~\ref{fig:metphase} shows the evolution of the
mean mass-weighted metallicity in various phases, along with the global
mean shown as the thick cyan line.  The global mean metallicity tracks
the diffuse IGM metallicity quite well down to $z\sim 1$, albeit higher by
around 0.2~dex.  This is largely because the diffuse phase contains most
of the baryons in the Universe until $z\sim 1$, where WHIM gas begins
to dominate.  The global metallicity is around 2-3\% solar at $z=2.5$
and reaches a present-day value of almost one-tenth solar.

It has been claimed that clusters represent a fair sample of the Universe
in terms of galaxy formation processes, because they have a stellar baryon
fraction similar to the global value~\citep[e.g.][]{ren98}.  However our
models do not support this idea~\citep[and neither do those of][]{cal04}.
Our predicted hot halo gas metallicity is well above the mean global
metallicity at all epochs, and does not evolve similarly.  This is
primarily because the interplay of outflows with cluster potentials allows
larger halos to retain and/or reaccrete more of their metals.  In other
words, galaxies are significantly affected by their environment not only
through structure formation processes, but also through outflow processes.

Stars gain a mean metallicity of $\ga 1/10$~solar very early on,
and in conjunction with a mass-metallicity relation whose slope is
established at early times~\citep{dav06}, it is not surprising to find
massive high-redshift galaxies with solar metallicity~\citep{sha04}.
The mean stellar metallicity rises steadily from around one-tenth
solar at $z=6$ to around one-half solar today.  This latter value
can be compared to observations, though there are difficulties in
determining stellar population metallicities owing to degenaracies with
age~\citep[e.g][]{tra00}.  \citet{fin03} found a mean stellar metallicity
of 0.6~solar in galaxy groups, which agrees nicely with our predicted
value, assuming that stars in galaxy groups are representative stars
everywhere in terms of their metallicity.

\begin{figure}
\vskip -0.4in
\setlength{\epsfxsize}{0.6\textwidth}
\centerline{\epsfbox{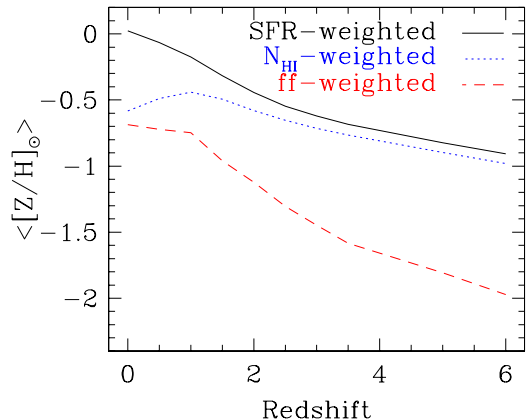}}
\vskip -3.0in
\caption{Mean gas metallicity weighted by star formation rate (solid),
\ion{H}{i} column density (dotted blue), and free-free emission (dashed red),
as a function of redshift.  The first represents the mean metallicity
of stars forming at a given redshift, the second corresponds roughly to
the mean metallicity in DLAs, and the final one corresponds roughly to
the X-ray weighted metallicity of intracluster gas.  }
\label{fig:metweight}
\end{figure}

Star-forming gas has a mean metallicity that evolves similarly to that
of stars, only about a factor of two higher.  Observational measures of
gas-phase metallicity in galaxies generally rely on emission line tracers
that arise in star forming regions.  Hence a more observationally relevant
question is:  What is the mean metallicity of stars forming today?
This can be calculated as the SFR-weighted metallicity.

Figure~\ref{fig:metweight} (solid line) shows the SFR-weighted metallicity
as a function of redshift.  This weighting is most closely related to the
\ion{H}{ii} region emission line metallicity used to determine, e.g., the
mass-metallicity relation in galaxies~\citep[e.g.][]{tre04}.  At early epochs,
the SFR-weighted metallicity is quite similar to the mean metallicity
in stars, but by $z=0$ it is roughly twice the stellar one.  At the
present epoch, our model predicts that stars are forming on average at
solar metallicity, although the average star has a metallicity half that.

The metallicity of DLAs provides another constraint on global metallicity
evolution.  Their \ion{H}{i} column density weighted metallicity is
seen to evolve slowly from one-tenth solar at $z\sim 3$ to one-fifth
solar at $z\sim 0.5$.  with some evidence for lower metallicities
at $z\ga 3$~\citep{per06}.  In contrast, models by \citet{pei99} and
\citet{som01}, which match DLA metallicities well at $z\ga 2$, predict
nearly solar metallicity at $z\sim 0.5$ for DLAs.

In order to compare our simulations to this data, we determine
an \ion{H}{i} column density-weighted mean metallicity, shown as
the dotted line in Figure~\ref{fig:metweight}.  This is obtained by
calculating the \ion{H}{i} ionization fraction assuming a \citet{haa01}
background as described in OD06, and multiplying the \ion{H}{i} density by
$\rho^{-1/3}$ to obtain a (scaled) column density for weighting.  In broad
agreement with data~\citep{per06}, the evolution is fairly mild from
$z\sim 3\rightarrow 0.5$, albeit somewhat higher than observed values.
In contrast with models mentioned above, our predicted DLA metallicity
actually drops from $z=1\rightarrow 0$, and is distinctly sub-solar by
the present epoch.

It is worth noting that our computation of the \ion{H}{i} fraction assumes
optically thin gas; this is clearly an incorrect assumption in the highly
dense regions that may be responsible for DLA absorption.  If we make
the extreme assumption that every star-forming gas particle is completely
neutral (even though it is likely that at least some portion of the ISM
is ionized), then the \ion{H}{i} column density-weighted metallicity
goes up by 0.2-0.4~dex at $z\la 2$.  Obviously a more careful analysis
is required; we leave this for future work.

The discrepancy between previous predictions and data has spurred
investigations of sub-DLAs as being responsible for holding highly
enriched \ion{H}{i} that may make up the shortfall.  Indeed, \citet{pro06}
and \citet{kul06} find highly enriched sub-DLAs, and suggest that they may
hold substantial amounts of metals, as discussed in \S\ref{sec:missing}.
However, while their metallicities are quite high, this is offset
by their lower column densities, so it is unclear whether they raise
the global \ion{H}{i} metallicity by a large amount.  From our models'
perspective, there is at most a minor discrepancy to explain: we predict
mild evolution and significantly sub-solar metallicities for \ion{H}{i}
gas even at low redshifts, consistent with observations, though the
somewhat higher metallicities overall predicted by our models could
certainly accomodate some contribution from highly enriched sub-DLAs.

A tight constraint for the metallicity of hot gas at $z\approx 0$
is provided by intracluster media, which show a relatively uniform
metallicity of around one-third solar.  Effectively, this is an X-ray
emission weighted metallicity.  Figure~\ref{fig:metweight} (dashed line)
shows the metallicity of gas with $T>10^6$~K weighted by its free-free
emission, i.e. $\rho^2 T^{0.5}$.  The value at $z=0$ is around one-fifth
solar, which is slightly below that seen in clusters.  Note however that
our simulation volume does not contain any actual clusters (our largest
halo has a virial temperature of around 2~keV), and also free-free
emission is a subdominant contribution to the emission at $T\la 1$~keV
where metal lines are prevalent.  So the modest disagreement, while
intriguing, should be regarded as preliminary.  It is still worth noting
that we predict little evolution in the emission-weighted metallicity
of hot gas to $z\sim 1$~\citep[in qualitative agreement with data,
e.g.][]{ros04}, but at higher redshifts this metallicity drops rapidly.

Shocked IGM metallicity remains relatively constant between one-hundredth
and one-fiftieth solar, and at $z\la 4$ it is very similar to the diffuse
phase metallicity.  The metallicity at $z\approx 0$ in this phase may be
tested using \ion{O}{vi} absorbers at low redshift~\citep[e.g.][]{tri06}.
Comparisons with simulations suggest that a typical metallicity of
one-tenth solar in \ion{O}{vi}-absorbing gas at low-$z$ would give rise
to a line frequency as observed~\citep{cen01,che03}.  These studies
pasted on a metal distribution to hydro simulations post facto.
Since our simulations produce a lower mean metallicity than inferred by
such studies, it bears testing whether our simulations can reproduce
the observed \ion{O}{vi} frequency directly from the simulated metal
distribution.  We are doing this comparison now.

In summary, our models show broad agreement with the observed metallicity
evolution as traced by stars, DLAs, and intracluster gas.  Clearly more
careful comparisons are warranted, but even these comparisons illustrate
the power of metal observations in constraining outflow models, and in
turn how such models can be used to coherently assemble information from
a variety of objects and environments.

\section{Metals in Phase Space}\label{sec:methist}

\subsection{Evolution in Phase Space}\label{sec:phasespace}

\begin{figure}
\vskip -0.5in
\setlength{\epsfxsize}{0.6\textwidth}
\centerline{\epsfbox{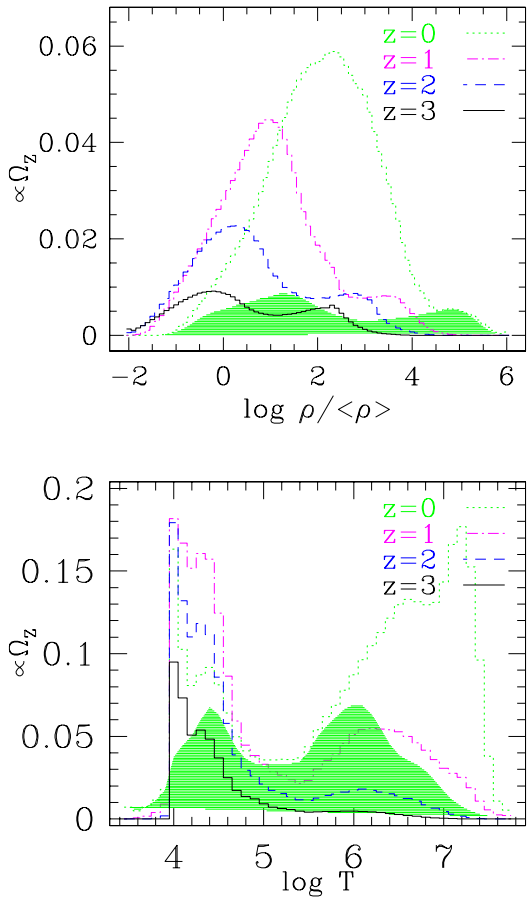}}
\vskip -0.5in
\caption{Histogram of metal mass in gas as a function of overdensity
(top panel), and temperature (bottom) at $z=3,2,1,0$.  The area under
the histogram is proportional to the total metal mass at that redshift.
Shaded regions show the $z=0$ histogram for $T<10^{4.5}$~K (top) and
for $\rho/\bar\rho<100$ (bottom).  The vertical axis values are arbitrary,
but are proportional to the total metals mass $\Omega_Z$.
}
\label{fig:methist}
\end{figure}

The evolution of metals cosmic phase space, i.e. overdensity versus
temperature, gives insights into galaxy formation processes.  Metals are
driven to lower densities through outflows, and to higher densities
through gravitational collapse.  They are heated by gravitational
infall and outflow-induced shocks, and cooled via radiative processes
and Hubble expansion.  The path of metals in phase space over time is a
key prediction of hierarchical galaxy formation models with outflows.
Here we describe some general trends of the path of cosmic metals in
phase space.  We focus here on metals in gaseous form, leaving aside
the metals locked into stars.

Figure~\ref{fig:methist} shows histograms of metal mass as a function of
overdensity (top panel) and temperature (bottom), at $z=3,2,1$ and 0.
The overdensity here is taken from the particle densities, hence it
is smoothed on the scale of the SPH kernel, which goes from hundreds
of kpc in underdense regions to sub-kpc scales in the densest regions;
it is the essentially the minimum scale over which density fluctuations
can be modeled with this SPH-based code.  At all redshifts, metals show
a characteristic bimodality in overdensity, with the diffuse peak holding
the majority of metals.  At $z=3$, metals are prominent in the diffuse IGM
with $\rho/\bar\rho\la 1$, but are also quite evident in a rather narrow
peak at $\rho/\bar\rho\sim 200$ that likely arises because outflows stall
as they intercept a dense IGM just outside halos.  At $z=2$, the diffuse
IGM increasingly dominates the gas metal budget, but the peak has shifted
towards higher overdensities.  The trend continues at $z=1$, and by by
$z=0$ the peak of metals has shifted to fairly sizeable overdensities,
around $\rho/\bar\rho\sim 200$, and metals are now seen in very dense gas.

The main trend from this is that metals shift to more overdense regions
with time, because large halos are able to retain more of their metals.
Note that in our simulations this is {\it not} because outflows escape
more easily from small halos.  Recall that in our momentum-driven
wind model, the outflow velocity scales with the velocity dispersion.
Hence all halos, to first order, lose a fixed amount of their ejected
material.  However, since the mass loading factor scales inversely with the
velocity dispersion, smaller halos eject a greater fraction of their
star-forming gas, causing them to lose more of their total metal mass.
Another factor is that, as star formation shuts off in the more massive
halos, the outflows also subside, and the earlier-forming massive halos
have more time to reaccrete their metals.

The bottom panel shows the histogram with respect to temperature.  Again,
a sort of bimodality exists, although it is not obvious until $z\la 2$.
Very few metals have $T<10^4$K; although the underdense photoionized IGM
extends to lower temperatures, the metals are restricted to the denser
regions, and have a relatively small filling factor~(OD06).  Note that the
cutoff at high temperatures is an artifact of our small simulation volume,
whose largest halo at $z=0$ has a virial temperature of $\approx 2$~keV.
While the amount of metals increases at all temperatures, the most notable
increase occurs at $T\ga 10^6$~K.  This owes primarily to the growth of
large enriched potential wells from surrounding enriched WHIM gas, which
is then unable to cool their gas efficiently~\citep{ker05}.  However, this
does not imply that cluster metallicities should evolve significantly,
because as seen in Figure~\ref{fig:metphase} this evolution is driven
mainly by an increase in the baryon fraction in the halo gas phase,
while its metallicity remains fairly constant.

The bimodality in both the density and temperature histograms
suggests that the peaks may be straightforwardly related.  But this
is not so.  To illustrate this, the shaded region in the top panel
of Figure~\ref{fig:methist} shows the low-temperature ($T<10^{4.5}$K)
portion of the $z=0$ density histogram, while the shaded region in the
bottom panel shows the low-density ($\rho/\bar\rho<100$) portion of the
$z=0$ temperature historgram.  This shows that the high-overdensity peak
is comprised primarily of cooler gas, while the major peak at moderate
overdensities is mostly shocked gas.  In the temperature histogram, the
low density gas dominates at $10^{4.5}\la T\la 10^6$~K (corresponding
to WHIM gas), while at $T\ga 10^6$~K and $T\la 10^{4.5}$~K one finds
higher-density gas.

In summary, gas-phase metals move into hotter, denser regions in the
Universe with time, owing to the interplay between early outflows,
gravitational reaccretion, and heating by structure formation.  At $z=2$,
the peak of metals is around the cosmic mean density, in diffuse
photoionized gas.  By $z=0$, the metals have recollapsed into regions more
than 100 times the mean density, with many of the metals shock heated
to $T>10^5$~K.  Tracing this evolution directly with observations would
provide a stringent test for cosmic enrichment models.

\subsection{IGM Metal Tracers}\label{sec:tracers}

How does one observationally catalog metals in various baryonic phases?
As discussed in \S\ref{sec:metphase}, metals in ISM gas are probably most
straightforward to catalog as they are visible in emission lines, though
significant calibration issues remain~\citep[e.g.][]{ell05}.  Stars are
also fairly straightforward to count, although the age-metallicity
degeneracy requires spectroscopy to overcome.  Metals in hot gas may be
seen in X-rays, which can in principle pick up a significant fraction of
metals at $z=0$ (Figure~\ref{fig:methist}), but this is limited to low
redshift systems having $T\ga 10^{6}$~K where the Galactic foregrounds are
not overwhelming.  The remainder of metals are probably most effectively
traced through absorption line spectroscopy against background sources
such as quasars, with the most commonly utilized tracers being ultraviolet
lines \ion{C}{iv} and \ion{O}{vi}.  In this section we examine how such
metal ions can trace the evolution of metals in phase space.

\begin{figure}
\setlength{\epsfxsize}{0.55\textwidth}
\centerline{\epsfbox{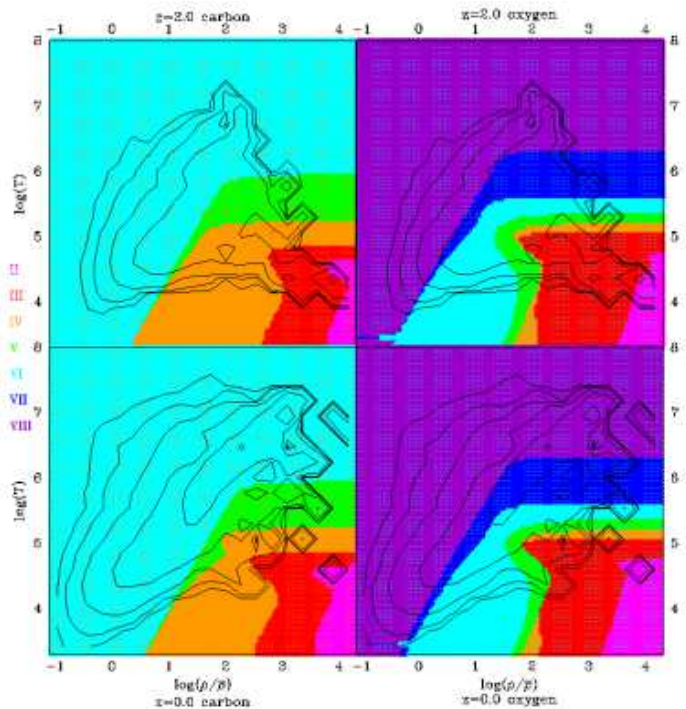}}
\vskip -0.2in
\caption{Phase space diagrams showing the metal mass (contours) and
regions of phase space where the ions shown are the dominant state of
carbon (left panels) and oxygen (right), at $z=2$ (top panels) and $z=0$
(bottom).  Exceptions are \ion{C}{iv}, \ion{O}{v}, and \ion{O}{vi}, which
are shown when their ionization fraction is above 8\% (or else they would
not show up at all).  Ionization states are colour coded as indicated to the left
of the plot.
}
\label{fig:metrhoT}
\end{figure}

Figure~\ref{fig:metrhoT} shows a phase space diagram for gas from our
simulations, with contours showing metal mass at $z=2$ (upper panels)
and $z=0$ (lower).  The evolution of metals in phase space as described
in the previous section is evident here, as metals move to hotter, denser
gas from $z=2\rightarrow 0$.  At low redshifts some metals do penetrate
well into the voids, affording at least in principle the ability to
trace low-density photoionized IGM gas with metal absorption lines,
although the ionization levels are quite high.

The coloured bands show regions in phase space where a given ion is the
dominant ionization state for that metal.  Left panels show carbon,
right panels show oxygen, and the colouring goes from singly-ionized
(in the lower right area) to highly ionized (in the upper left area).
For visibility, \ion{C}{iv}, \ion{O}{v}, and \ion{O}{vi} are always
shown when they have ionization fractions above 8\%; this is done because
those species are never the dominant ionization state, and hence would
otherwise not show up on the plot.

The key point to note on this figure is the overlap region between a
given ion and the metal mass contours.  This corresponds to the range of
densities and temperatures that are best traced by that ion.  Of course,
it is often possible to detect multiple ions for a given system,
and the detectibility in practice is influenced by other factors such
as the oscillator strength and the efficiency of detectors in a given
wavelength regime.  Hence this plot is only intended to be illustrative
of broad trends.  Generally, the bands show a characteristic pattern
on the plot, with a diagonal rise from low temperatures and densities
owing to photoionization, and then a plateau at a particular temperature
corresponding to the collisional ionization maximum for that given ion.
Note that this plot was computed using CLOUDY~\citep{fer98} with a
\citet{haa01} ionizing background and assuming ionization equilibrium.
The photoionized region is somewhat sensitive to our assumed ionizing
background, but the collisional ionization plateau is not.

From this plot it is straightforward to see why \ion{C}{iv}
(1548,1551\AA) is a popular absorption feature for tracing IGM metals,
being an easily-identifiable doublet at observationally-accessible
wavelengths.  \ion{C}{iv} overlaps the enriched region of phase space
from $\rho/\bar\rho\sim 10-1000$ with $T\la 10^5$~K at $z=2$.  As OD06
emphasize, \ion{C}{iv} becomes a progressively better tracer of bulk IGM
metals to higher redshifts.  Conversely, to lower redshifts it shifts
to higher overdensities and becomes a poorer tracer of IGM metals.
\ion{C}{vi} (33.7\AA) is more optimal for tracing photoionized IGM metals
and \ion{C}{v} (40.3\AA) nicely tracks the bulk of metals (and gas) in
the WHIM regime, but both lie in the observationally challenging soft
X-ray regime.  At $z=2$, \ion{C}{iii} (977\AA) tracks the metals in cool
gas around galaxies, but by $z=0$ it is predicted to be uncommon because
there are few metals left in such dense regions, as it has mostly been
consumed in star formation or shock-heated to higher temperatures.
\ion{C}{ii} (1335\AA) traces gas in the inner regions of galactic
halos, which likely involves multi-phase gas clumps for which it becomes
difficult to interpret line ratios.  Hence until significant advances are
made in soft X-ray spectrographs, carbon is in practice only effective
for tracing IGM metals at $z\ga 2$.

Oxygen has the advantage of being a highly abundant element, and if
starburst ejectae from Type II supernovae are primarily responsible
for enriching the IGM, may be alpha-enhanced relative to carbon.
Furthermore, a wide range of oxygen lines are observationally
accessible and fairly strong.  The most common line used to see IGM
metals is \ion{O}{vi} (1032,1037\AA), which traces a wide range of
environments from photoionized gas around the cosmic mean up to WHIM
gas at overdensities around a thousand.  Indeed, disentangling the
photoionized \ion{O}{vi} absorbers versus the collisionally ionized ones
is challenging~\citep{tri01,tri06}.  At $z\ga 2.5$ this line starts
to become overwhelmed by the thickening \lya\ forest~\citep{sim04}.
\ion{O}{vii} (21.6\AA) and \ion{O}{viii} (19\AA) trace the majority of
WHIM metals~\citep[e.g.][]{che03}, as well as covering the bulk of the
photoionized metals at lower redshifts.  However, being that these are
X-ray lines, the photoionized regions have too low densities to provide
detectable absorption at current sensitivities, and the warm-hot systems
are only seen with great effort~\citep[][though this detection is
controversial; see \citep{kaa06} and \citep{ras06}]{nic05}.  Lower ionization lines
like \ion{O}{v} (630\AA) and \ion{O}{iv} (787\AA) are interesting;
although they do not dominate over a wide swath of phase space, it
may be possible to detect them with UV instruments in conjunction with
\ion{O}{vi} at moderate redshifts ($z\ga 1$), enabling detailed studies
of the physical conditions and metallicity of $T\sim 10^5$~K absorbers.
\ion{O}{iii} (702\AA) traces denser, colder gas which at $z\ga 2$ contains
substantial amounts of metals, and \ion{O}{ii}, like \ion{C}{ii}, is
limited to the regions immediately around star-forming gas.

In the future it may be possible to detect IGM metal lines in emission,
though this will require a new UV-capable instrument such as the proposed
Baryonic Structure Probe~\citep{sem05}.  The most natural ions to target
from a practical standpoint are as usual \ion{O}{vi} and \ion{C}{iv},
because UV detectors are efficient down to $\sim 1000$\AA.  However,
it would be highly desireable to have detector capabilities extending
down to $\sim 700$\AA, so that a suite of \ion{O}{iii}-\ion{O}{vi} can
be seen in relatively nearby large-scale structure where the emission
is not hopelessly dimmed by surface brightness effects.  There is also
development underway for X-ray imaging with the Diffuse Intergalactic
Oxygen Surveyor~\citep{oha06}.  Imaging offers the opportunity in 
principle to directly trace the filamentary topology of WHIM gas,
providing another interesting test of models.

In summary, performing a full inventory of metals in intergalactic
gas will require next-generation UV and X-ray spectrographs capable of
detecting species that sample a wide range of ionization conditions.
To this end, oxygen seems more promising than carbon, as \ion{O}{iii}
to \ion{O}{vi} are all accessible in the UV, and \ion{O}{vii} and
\ion{O}{viii} are relatively strong X-ray lines, with oxygen being
a highly abundant element.  With statistical information on at least
several of these lines for a large number of systems, it will become
possible to observationally characterize the metal distribution (along
with the density and temperature distribution) of IGM gas, and directly
test model predictions.  For the time being, we are mostly limited to studying
single ions in a small number of systems, and inferring the total metal
mass from indirectly determined ionization conditions or by comparing
with models such as the ones presented here.

\section{Conclusions}\label{sec:summary}

Using cosmological hydrodynamic simulations of galaxy formation including
galactic outflows, we have studied the distribution and evolution of
metals in various baryonic phases in the Universe.  Our main conclusions
are summarized as follows:

\begin{itemize}

\item Metals are present in substantial amounts in a wide range of
environments from the earlier epochs studied ($z\sim 6$) until the
present.  Diffuse IGM gas dominates the metal budget early on, while
stars and shocked gas hold most of the metals today.

\item The missing metals (outside of galaxies) at $z\sim 2.5$ are mainly
in intergalactic gas that is mildly hotter than pure photoionization
temperatures, having $10^4\la T\la 10^5$~K.  However, this gas is not
confined to halos, but is spread across a wide range of overdensities,
with a peak around the cosmic mean density.

\item Comparisons of our simulation to observations broadly show agreement
in the amount of metals in various baryonic phases, and show that metals
are spread relatively evenly across all phases.

\item The global mean metallicity tracks about 50\% higher than the
diffuse IGM metallicity from $z=6\rightarrow 1$, after which it continues
increasing to about one-tenth solar at the present epoch.  Hot halo gas,
which is predicted to have a luminosity-weighted metallicity of around
one-fifth solar today, does not track the global metallicity, suggesting
that clusters are unlikely to be representative samples of the universe
at large.  This conclusions comes with a significant caveat, however,
as our simulation volume is too small to produce actual clusters, so
improved simulations are needed to confirm this.

\item Stars and star-forming gas quickly obtain a mean metallicity
of $\ga 0.1Z_\odot$ solar at $z\sim 6$, indicating early enrichment
in galaxies.  Today stars have a mean metallicity of one-half solar,
while stars just forming now have typically solar metallicity.

\item Our models predict a global \ion{H}{i} column density-weighted
metallicity that increases slowly with redshift from $z=6\rightarrow 1$,
and then actually decreases to $z=0$, never exceeding around one-third
solar.  This is broadly consistent with observations of DLA metallicities,
and does not require a substantial contribution from sub-DLAs.

\item Metals generally move from lower temperature and more diffuse
regions at early epochs towards hotter, more overdense regions at
low redshifts.  This occurs because early blowout is very efficient
at enriching diffuse gas, while at lower redshift outflows subside and
large potential wells reaccrete gas.

\item Oxygen, particularly \ion{O}{iii}--\ion{O}{vii}, provides the
most effective tracer of metals across a wide range of cosmic phase
space at $z\la 2$, but a full metal accounting in the IGM will probably require
next-generation instruments.

\end{itemize}

Obtaining a complete census of metals in the Universe will require a
concerted effort in studying a variety of systems and environments at
various epochs.  Still, there is a realistic possibility of providing a
full census of metals in the Universe in the foreseeable future.  This is
certainly a formidable challenge for upcoming facilities, but the payoff
is large.  By comparing such observations with cosmological simulations
of outflows that track the interplay between various baryonic phases in
mass, metals, and energy, it will be possible to stringently constrain
models of galaxy formation and feedback.  This will dramatically advance
our understanding of how galaxies and intergalactic gas interact in
the cycle of accretion and feedback, and provide for the first time a
complete view of how baryons evolve in cosmic phase space from the dark
ages until the present.

 \section*{Acknowledgements}
We thank MPIA and H.-W. Rix for their hospitality during the writing of
this paper.  The simulation used here was run on grendel, our department's
Beowulf system at the University of Arizona.  Support for this work,
part of the Spitzer Space Telescope Theoretical Research Program,
was provided by NASA through a contract issued by the Jet Propulsion
Laboratory, California Institute of Technology under a contract with NASA.
Support for this work was also provided by NASA through grant number
HST-AR-10647.01 from the SPACE TELESCOPE SCIENCE INSTITUTE, which is
operated by AURA, Inc. under NASA contract NAS5-26555.

\end{document}